\documentclass[fleqn,twoside]{article}
\usepackage{espcrc2}

\usepackage{graphicx}
\usepackage[figuresright]{rotating}

\newcommand{\AmS}{{\protect\the\textfont2
  A\kern-.1667em\lower.5ex\hbox{M}\kern-.125emS}}

\title{Jet measurements at D\O\ using a $k_T$ algorithm}

\author{V.\ Daniel Elvira\address[MCSD]{Fermi National Accelerator Laboratory,
P.\O.\ Box 500, Batavia, IL, 60510-500, USA.} for the D\O\ Collaboration.}
       
\begin{document}

\begin{abstract}
D\O\ has implemented and calibrated a $k_{\perp}$ jet algorithm for the first
time in a $p\overline{p}$ collider.  We present two results based on
$1992$-$1996$ data which were
recently published: the subjet multiplicity in quark and gluon jets and the 
central inclusive jet cross section. The measured ratio between subjet 
multiplicities in gluon and quark jets is consistent with theoretical 
predictions and previous experimental values. 
NLO pQCD predictions of the $k_{\perp}$
inclusive jet cross section agree with the D\O\ measurement, although
marginally in the low $p_{\mathrm T}$ range.
We also present a preliminary measurement of thrust cross sections, which
indicates the need to include higher than $\alpha_{\mathrm s}^{3}$ terms and
resumation in the theoretical calculations.
\vspace{1pc}
\end{abstract}

\maketitle

\section{INTRODUCTION}
\label{sec:intro}

Until recently,  only cone algorithms were used to reconstruct jets
in $p\overline{p}$~\cite{snowmass} colliders.
The cone algorithm used to reconstruct the $1992$-$1996$ Tevatron 
data~\cite{d0_jets_prd,cdf_jets_prd} presents several shortcomings:

\begin{itemize}
\item An arbitrary procedure must be implemented to split and
merge overlapping calorimeter cones.
\item An ad-hoc parameter, $R_{\mathrm{sep}}$~\cite{rsep}, is required to 
accommodate the differences between jet definitions at the parton and 
detector levels.
\item Improved theoretical predictions calculated at the
next-to-next-to-leading-order (NNLO) in pQCD are not infrared safe,
but exhibit sensitivity to soft radiation.
\end{itemize}

A second class of jet algorithms, free of these problems, was 
developed during the past decade~\cite{catani93,catani92,ellis93}.  
These recombination
algorithms successively merge pairs of nearby objects (partons,
particles, or calorimeter towers) in order of increasing relative
transverse momentum.  A single parameter, $D$, which approximately
characterizes the size of the resulting jets, determines when this
merging stops.  No further splitting or merging is involved because each
object is uniquely assigned to a jet.  There is no need to introduce
any ad-hoc parameter, because the same algorithm is applied at the
theoretical and experimental level.  Furthermore, by design,
clustering algorithms are infrared and collinear safe to all orders of
calculation.

The D\O\ Collaboration has 
implemented a $k_{\perp}$ algorithm to reconstruct jets from data taken 
during the $1992$-$1996$ collider run. This paper is a 
review of the associated measurements recently performed by
the D\O\ experiment~\cite{rob,seb}.

\section{THE RUN I D\O\ DETECTOR}

D\O\ is a multipurpose detector designed to study $p\bar{p}$
collisions at the Fermilab Tevatron Collider.
A full description of the Run I D\O\ detector can be found in Ref.~\cite{d0}.
The primary detector components for jet measurements at D\O\
are the calorimeters, which use liquid-argon as the active medium and 
uranium as the absorber.
The D\O\ calorimeters provide full solid angle coverage and particle
containment (except for neutrinos or high $p_{\mathrm T}$ muons), as well as
linearity of response with energy and compensation ($e/\pi$ response ratio is
less than 1.05).
Transverse segmentation is $\Delta \eta \times \Delta \phi = 0.1 \times 0.1$,
with $\eta = - \mathrm{ln} \left[ \mathrm{tan} (\theta / 2) \right]$.
The single particle energy resolutions for electrons (e) and pions ($\pi$), 
measured from test beam data, are approximately 15$\%$ and 50$\%$ respectively.

\section{$\lowercase{k}_{\perp}$ JET ALGORITHM}
\label{sec:alg}

The D\O\ $k_{\perp}$ jet algorithm~\cite{rob} starts with a list of energy
pre-clusters, formed from calorimeter cells, final state 
partons or particles. The angular separation between pre-clusters is
$\Delta{\cal R} = \sqrt{ \Delta \eta^2 + \Delta \phi^2} > 0.2$.

\begin{figure}[htbp]
\includegraphics[width=7cm]{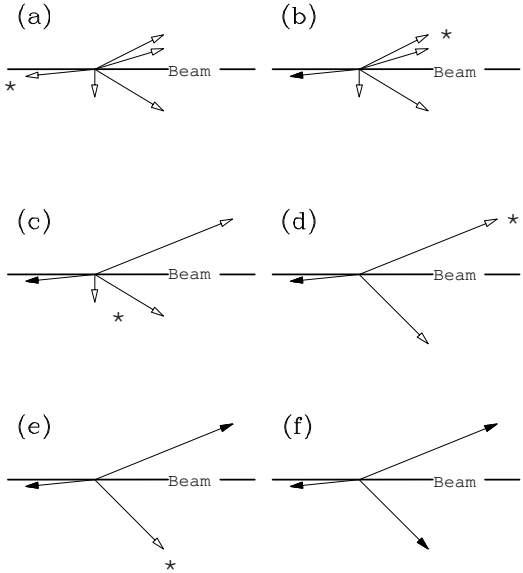}
\vskip-1cm
\caption{Example of the $k_{\perp}$ merging mechanism on a 
hard $p\overline{p}$ event.
(a) The particles represented by arrows comprise a list of objects.
(b-f) Solid arrows represent
the final jets reconstructed by the $k_{\perp}$ algorithm,
and open arrows represent objects not yet assigned to jets.
The five diagrams show successive iterations of the algorithm.
In each diagram, an object is labeled as a jet if it is
well separated from all other objects. Two objects are merged if 
they have small relative $k_{\perp}$.
The asterisk labels the relevant objects at each step.}
\label{fig:kt_example}
\end{figure}

The jet reconstruction is performed in three steps:

1. For each object $i$ in the list, the algorithm defines
$d_{\mathrm{ii}} = p_{\mathrm{T,i}}^2$, where $p_{\mathrm T}$ is 
the momentum transverse to the beam.
For each pair $(i,j)$ of objects, it also defines
$d_{\mathrm{ij}} = \mathrm{min}(p_{\mathrm{T,i}}^2,p_{\mathrm{T,j}}^2) 
\frac{ \Delta{\cal R}_{\mathrm{ij}}^2} {D^2}$,
where $D$ is a resolution parameter.

2. If the minimum of all possible $d_{\mathrm{ii}}$ and $d_{\mathrm{ij}}$ is a 
$d_{\mathrm{ij}}$, the algorithm replaces objects $i$ and $j$ by 
their 4-vector sum before going back to step~1.
If the minimum is a $d_{\mathrm{ii}}$, then $i$ is defined as a jet and
removed from the list of objects.

3. While there is an object left in the list, the algorithm returns to step~1.

The final product of this process is a list of jets, separated by 
$\Delta{\cal R} > D$ from each other.
Subjets may be defined by re-running the $k_{\perp}$ algorithm
from a list of pre-clusters in a given jet.
Pairs of objects with the smallest $d_{\mathrm{ij}}$ are merged successively 
until all remaining $d_{\mathrm{ij}}$ satisfy 
$d_{\mathrm{ij}} > y_{\mathrm{cut}} p_{\mathrm{Tjet}}^{2}$. 
The resolved objects
are called subjets, and the number of subjets within the jet 
is the subjet multiplicity $M$.
For $y_{\mathrm{cut}}$=$1$, the entire jet consists of a single subjet 
($M$=$1$).
As $y_{\mathrm{cut}}$ decreases, the subjet multiplicity increases, until
every pre-cluster becomes resolved as a separate subjet
in the limit $y_{\mathrm{cut}} \rightarrow 0$.

\section{CALIBRATION OF JET MOMENTUM}
\label{sec:jes}

The uncertainty in the jet energy or momentum is the dominant error
in almost every jet measurement at a hadron collider. The jet momentum
calibration is described in Ref.~\cite{rob,jes}.
The calibration at D\O\ accounts for detector effects like response, noise,
and signal pile-up from previous crossings. It also removes the
underlying event formed by the remnant soft partons (u.e.), and
the contribution of multiple $p\overline{p}$ interactions.
These corrections enter a relation
between the momentum of a jet measured in the calorimeter 
$p_{\mathrm{jet}}^{\mathrm{meas}}$
and the ``true'' jet momentum $p_{\mathrm{jet}}^{\mathrm{true}}$:

\begin{equation}
p_{\mathrm{jet}}^{\mathrm{true}} = 
\frac{p_{\mathrm{jet}}^{\mathrm{meas}} - 
p_O(\eta^{\mathrm{jet}},{\cal L},p_{\mathrm T}^{\mathrm{jet}})}
{R_{\mathrm{jet}}(\eta^{\mathrm{jet}},p^{\mathrm{jet}})}
\label{eq:jes}
\end{equation}

where the offset term $p_O$ corrects for u.e., noise, pile-up, and
multiple interactions, while $R_{\mathrm{jet}}$ corrects for the response of
the calorimeter to jets.
The true jet momentum is defined as the particle level jet momentum.
A particle level jet is reconstructed from the final state particles, after
hadronization but before interaction with the calorimeter material.
The calibration procedure follows closely
that of the calibration of the fixed-cone jet algorithm~\cite{jes}.

The fractional momentum resolution for $k_{\perp}$ jets ($D$=$1$)
is determined from the measured $p_{\mathrm{T}}$ imbalance in dijet events.
At 100 (400)~GeV, the fractional resolution is 
0.061$\pm$0.006(0.039$\pm$0.003). 
Within statistical and systematic 
uncertainties, there is not a significant difference between energy
resolutions associated with 
$k_{\perp}$ ($D$=$1$) and cone ${\cal R}$=$0.7$ jets.

\section{PHYSICS RESULTS}

D\O\ has performed
a number of measurements using the $k_{\perp}$ jet algorithm.
These include a study of the structure of quark and gluon jets~\cite{rob}, and
measurements of the central ($|\eta|<0.5$) inclusive jet cross  
section~\cite{seb}, and thrust distributions.

\subsection{Subjet multiplicities}

In LO QCD, 
the fraction of final state jets which are gluons decreases
with $x \sim p_{\mathrm T} / \sqrt{s}$, 
the momentum fraction of initial state partons within the proton.
For fixed $p_{\mathrm T}$, the gluon jet fraction decreases when 
$\sqrt{s}$ is decreased from 1800~GeV to 630~GeV.
We select gluon and quark enriched samples 
with identical cuts in events at $\sqrt{s} = 1800$ and 630~GeV
to reduce experimental biases and systematic effects.
Of the two highest $p_{\mathrm T}$ jets in the event, 
we select 
$k_{\perp} (D=0.5)$ jets with 
$55 < p_{\mathrm T} < 100$~GeV and $|\eta| < 0.5$.

There is a simple method to extract a measurement
of quark and gluon jets on a statistical basis.
If $M$ is the subjet multiplicity 
in a mixed sample of quark and gluon jets,
it may be written as a linear combination of subjet multiplicity
in gluon and quark jets:
\begin{equation}
M=fM_{\mathrm g}+(1-f)M_{\mathrm q}
\label{eq:m}
\end{equation}
The coefficients are the fractions of gluon and
quark jets in the sample, $f$ and $(1-f)$, respectively. 
Consider Eq. (\ref{eq:m}) for two
similar samples of jets at $\sqrt{s} = 1800$ and 630~GeV,
assuming $M_{\mathrm g}$ and $M_{\mathrm q}$
are independent of $\sqrt{s}$.
The solutions are

\begin{equation}
M_{\mathrm q}=\frac{f^{1800}M^{630}-f^{630}M^{1800}}{f^{1800}-f^{630}}  
\label{eq:mq}
\end{equation}

\begin{equation}
M_{\mathrm g}=\frac{\left( 1-f^{630}\right) M^{1800}-\left( 1-f^{1800}\right) 
M^{630}%
}{f^{1800}-f^{630}}  \label{eq:mg}
\end{equation}

where $M^{1800}$ and $M^{630}$ are the
experimental measurements in the mixed jet samples at  
$\sqrt{s} = 1800$ and 630~GeV,
and $f^{1800}$ and $f^{630}$ are the gluon jet fractions in the two samples.
The method relies on knowledge of the two gluon jet fractions, which
are extracted from the {\sc herwig} 5.9\cite{hw} Monte Carlo event 
generator and used in Eqs. (\ref{eq:mq}-\ref{eq:mg}).

\begin{figure}[htbp]
\vskip-5mm
\includegraphics[width=7cm]{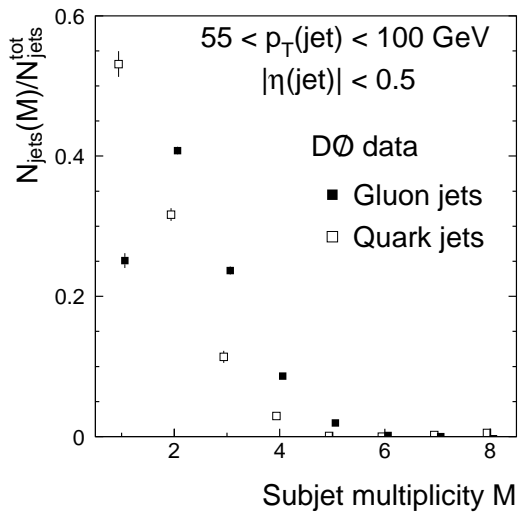}
\vskip-1cm
\caption{
Corrected subjet multiplicity for gluon and quark jets, 
extracted from D\O\ data.}
\label{fig:d0cqga}
\vskip-1cm
\end{figure}

\begin{figure}[htbp]
\includegraphics[width=7cm]{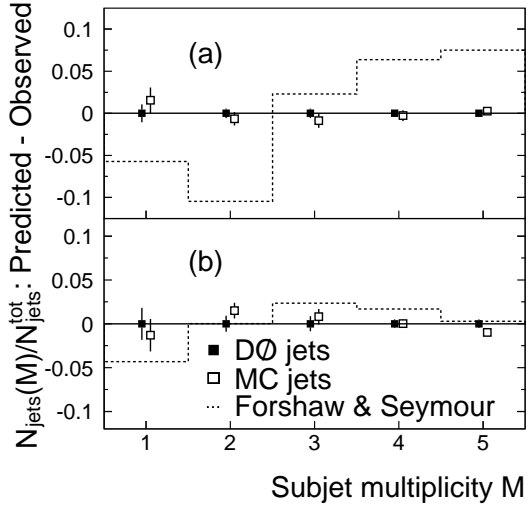}
\vskip-1cm
\caption{
The subjet multiplicity in (a) gluon and 
(b) quark jets, for D\O\ data, for the {\sc herwig} Monte Carlo, and 
resummed predictions.}
\label{fig:d0cqgb}
\vskip-5mm
\end{figure}

Figure~\ref{fig:d0cqga} shows the average 
subjet multiplicity ($y_{\mathrm{cut}} = 10^{-3}$) for quark 
and gluon jets. $M_{\mathrm g}$ is 
significantly larger for gluon jets than for quark jets.
The gluon jet fractions are the dominant
source of systematic error. 
We also compute the ratio
$R = \frac{\langle M_g \rangle - 1} {\langle M_q \rangle - 1}
= 1.84 \pm 0.15 \mathrm{(stat)} ^{+0.22}_{-0.18} \mathrm{(sys)}$.
Figure~\ref{fig:d0cqgb} shows a comparison between the
ratio measured by D\O\ , the {\sc herwig} 5.9 result of r=1.91,
the ALEPH\cite{aleph} value of r=1.7$\pm0.1$ ($e^{+}e^{-}$ annihilations at 
$\sqrt{M_{\mathrm Z}}=M_{\mathrm Z}$) , and the associated Monte Carlo 
and resummation prediction~\cite{sey_sub0}. Good agreement is observed. 
All of the experimental and theoretical values for $r$ 
are smaller than the naive QCD prediction value of 2.25 for the ratio 
of color charges.
This is because of higher-order radiation in QCD,
which tends to reduce the ratio from the naive value.

\subsection{Central inclusive jet cross section}

The inclusive jet cross section for $|\eta|<0.5$, 
$d^2\sigma/(dp_{\mathrm T} d\eta)$, 
was measured as $N/(\Delta\eta
\Delta p_{\mathrm T} \epsilon L)$, where $\Delta\eta$ and 
$\Delta p_{\mathrm T}$ are the
$\eta$ and $p_{\mathrm T}$ bin sizes, $N$ is the number of jets reconstructed
with the $k_{\perp}$ ($D$=$1$) algorithm in that
bin, $\epsilon$ is the overall efficiency for jet and event selection,
and $L$ represents the integrated luminosity of the data sample~\cite{seb}.

\begin{figure}[htbp]
\vskip-5mm
\includegraphics[width=8cm]{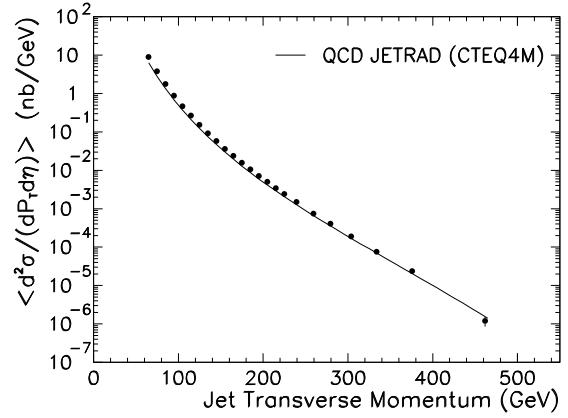}
\vskip-1cm
\caption{The central ($|\eta|<0.5$) inclusive jet cross 
section obtained with the $k_{\perp}$ algorithm at $\sqrt{s}=1.8$~TeV. 
Only statistical errors are included. The solid line shows a 
prediction from NLO pQCD.}
\label{fig:xsectiona}
\end{figure}

\begin{figure}[htbp]
\vskip-5mm
\includegraphics[width=8cm]{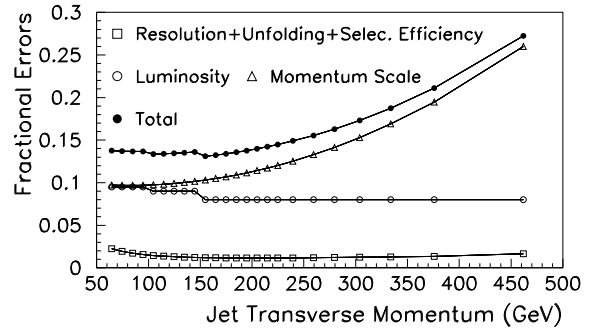}
\vskip-1cm
\caption{Fractional experimental uncertainties 
on the cross section.}
\label{fig:xsectionb}
\vskip-5mm
\end{figure}

The fully corrected cross section for $|\eta|<0.5$ is shown in
Fig.~\ref{fig:xsectiona}, along with the statistical uncertainties.
The systematic uncertainties include contributions from the jet and event
selection, unsmearing, luminosity, and the uncertainty in the momentum
scale, which dominates at all transverse momenta.  The fractional
uncertainties for the different components are plotted in
Fig.~\ref{fig:xsectionb} as a function of the jet transverse
momentum.

\begin{figure}[htbp]
\includegraphics[width=7.5cm]{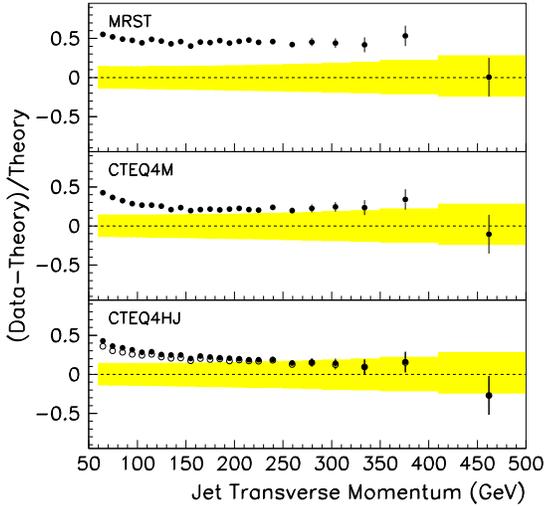}
\vskip-1cm
\caption{Difference between data and {\sc jetrad} pQCD, normalized to the
predictions. The shaded bands represent the total systematic uncertainty.
In the bottom plot a {\sc herwig} hadronization contribution has been 
added to the prediction (open circles).}
\label{fig:dtta}
\vskip-5mm
\end{figure}

The results are compared to the NLO pQCD prediction from {\sc
jetrad}~\cite{jetrad}, with the renormalization and factorization
scales set to $p_{\mathrm T}^{\mathrm{max}}/2$, where 
$p_{\mathrm T}^{\mathrm{max}}$ refers to
the $p_{\mathrm T}$ of the leading jet in an event. The comparisons are made
using parameterizations of the parton distribution functions (PDFs) of
the CTEQ~\cite{cteq} and MRST~\cite{mrs} families.
Figure~\ref{fig:dtta} shows the ratios of (data-theory)/theory.  The
predictions lie below the data by about $50\%$ at the lowest $p_{\mathrm T}$ 
and
by (10-20)$\%$ for $p_{\mathrm T}>200$~GeV.  To quantify the comparison in
Fig.~\ref{fig:dtta}, the fractional systematic uncertainties are
multiplied by the predicted cross section, and a $\chi^2$ comparison,
using the full correlation matrix, is carried out~\cite{d0_jets_prd}.
Though the agreement is reasonable ($\chi^2/\mathrm{dof}$ ranges from 
$1.56$ to $1.12$, the probabilities from $4$ to $31\%$), 
the differences in normalization and shape, especially at low 
$p_{\mathrm T}$, are quite large.  The
points at low $p_{\mathrm T}$ have the highest impact on the $\chi^2$.  If the
first four data points are not used in the $\chi^2$ comparison, the
probability increases from $29\%$ to $77\%$ when using the CTEQ4HJ
PDF.

\begin{figure}[htbp]
\vskip-5mm
\includegraphics[width=8cm]{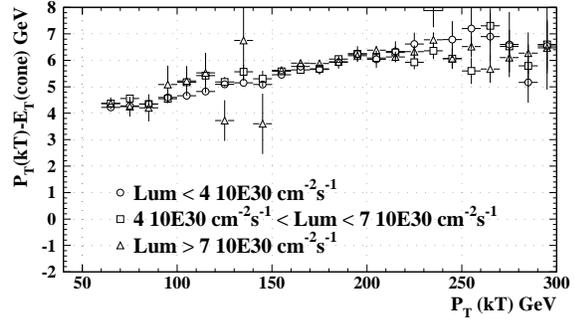}
\vskip-1cm
\caption{$p_{\mathrm T}$ of 
$k_{\perp}$ jets ($D$=$1$) minus $E_{\mathrm T}$ of the associated cone jets 
(${\cal R}$=$0.7$) for samples taken at different instantaneous luminosities.}
\label{fig:nikos}
\vskip-5mm
\end{figure}

The NLO predictions of the inclusive cross section for
$k_{\perp}$ ($D$=$1$) and cone jets 
(${\cal R}$=$0.7$, $R_{\mathrm{sep}}$=$1.3$)
in the same $|\eta|<0.5$ interval are within a few percent of each other in
the $p_{\mathrm T}$ range relevant in this analysis~\cite{sebtes}. 
The measured $k_{\perp}$ cross section, however, is $37\%$ ($16\%$) higher 
than the published cone algorithm~\cite{d0levan} at $60$
($200$)~GeV. This difference in the cross sections is consistent with
the measured difference in $p_{\mathrm T}$ for cone jets matched in 
$\eta$-$\phi$
space to $k_{\perp}$ jets, as shown in Fig.~\ref{fig:nikos}.
For the same energy clusters the $p_{\mathrm T}$ of 
$k_{\perp}$ jets ($D$=$1$) is higher than the 
$E_{\mathrm T}$ of associated cone jets (${\cal R}$=$0.7$).
The difference increases with jet $p_{\mathrm T}$,
from about 5~GeV (or $5 \%$) at $p_{\mathrm T} \approx 100$~GeV
to about 7~GeV (or $3 \%$) at $p_{\mathrm T} \approx 250$~GeV~\cite{rob}.
Fig.~\ref{fig:nikos} proves, however, that the energy difference does not
depend on the instantaneous luminosity associated with the sample. 
After offset subtraction, it is clear that $k_{\perp}$ jets are not 
contaminated by energy coming from pile-up, uranium noise, or 
multiple interactions.

\begin{figure}[htbp]
\vskip-5mm
\includegraphics[width=8cm]{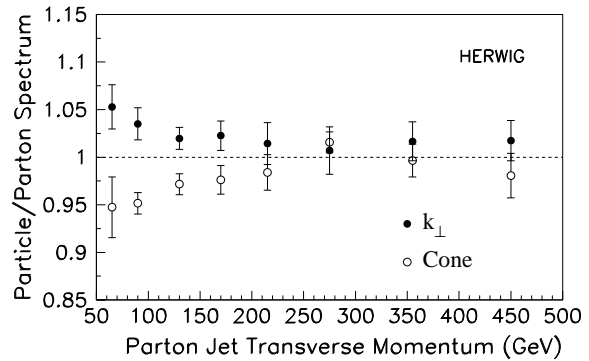}
\vskip-1cm
\caption{Ratio of particle-level over 
parton-level {\sc herwig} $p_{\mathrm T}$ spectra  for jets, 
as a function of the parton 
jet transverse momentum.}
\label{fig:dttb}
\vskip-5mm
\end{figure}

The effect of final-state hadronization on reconstructed energy, which
might account for the discrepancy between the observed cross section
using $k_{\perp}$ and the NLO predictions at low $p_{\mathrm T}$, and also 
for the
difference between the $k_{\perp}$ and cone results, was studied using {\sc
herwig} (version $5.9$) simulations.
Figure~\ref{fig:dttb} shows the ratio of
$p_{\mathrm T}$ spectra for particle-level to parton-level jets, for both the 
$k_{\perp}$ 
and cone algorithms.  Particle cone jets, reconstructed from final
state particles (after hadronization), have less $p_{\mathrm T}$ than the 
parton
jets (before hadronization), because of energy loss outside the cone.

In contrast, $k_{\perp}$ particle jets are more energetic than their
progenitors at the parton level, due to the merging of nearby partons
into a single particle jet.  Including the hadronization effect
derived from {\sc herwig} in the NLO {\sc jetrad} prediction improves
the $\chi^2$ probability from $29\%$ to $44\%$ ($31\%$ to $46\%$) when
using the CTEQ4HJ (MRST) PDF.  

We have also investigated the sensitivity of the measurement to the 
modeling of the background from
spectator partons through the use of minimum bias events, and found
that it has a small effect on the cross section: at low $p_{\mathrm T}$, where
the sensitivity is the largest, an increase of as much as $50\%$ in
the underlying event correction decreases the cross section by less
than $6\%$.

\subsection{Thrust cross sections}

Event shape variables have been extensively used in $e^{+}e^{-}$ and
$ep$ collider experiments to study the spatial distribution of
hadronic final states, to test the predictions of perturbative QCD,
and to extract a precise value of the coupling constant $\alpha_{\mathrm s}$.
Over the last few years, they have attracted considerable interest,
as they have proved to be a fruitful testing ground for
recent QCD developments like
resummation calculations and non-perturbative corrections.

There are several observables which characterize the
shape of an event. To be calculable by perturbation
theory, these quantities must be
infra-red safe, {\em i.e.} insensitive to the emission of soft
or collinear gluons. 
A widely used variable that meets this requirement is the thrust,
defined as
\begin{equation}
T = \mathrm{max}_{\hat{n}}\frac{\sum_{\mathrm i} \left|\vec{p}_{\mathrm i} 
\cdot \hat{n}\right|}{\sum_{\mathrm i} \left|\vec{p}_{\mathrm i}\right|}
\label{eq:Th}
\end{equation}
where the sum is over all partons, particles or calorimeter towers
in the event. The unit vector $\hat{n}$ that maximizes the ratio of the sums 
is called the thrust axis.
The values of thrust range from $T$=$0.5$ for a perfectly spherical
event, to $T$=$1$ for a pencil-like event, when all emitted particles are
collinear. In this latter case, the thrust axis lies along the direction of
the particles.

In most of the kinematic range, $e^{+}e^{-}$ and $ep$ collider 
experiments~\cite{epee} report good agreement of event shape distributions 
with $O(\alpha_{\mathrm s}^2)$ pQCD corrections to the lowest order 
QED diagram 
that governs the interaction.
Fixed order QCD calculations, however, fail when two widely
different energy scales are involved in the event, leading to the
appearance of large logarithmic terms at all orders in the perturbative
expansion~\cite{resum}. This happens in the limit of the two jet
back-to-back configuration, when $T\rightarrow 1$. This case is
handled by the resummation technique, which identifies the 
large logarithms in each order of
perturbation theory and sums their contributions to all orders.
DELPHI reports excellent agreement of thrust distributions
in $Z\rightarrow$ hadrons once resummation and hadronization corrections 
are added to the $O(\alpha_{\mathrm s}^{2})$ QCD prediction~\cite{delphit}.

In a hadron collider, it is convenient to introduce ``transverse thrust'',
$T^{\mathrm T}$, a Lorenz invariant quantity under 
$z$-boosts, which is obtained
from Eq.~\ref{eq:Th} in terms of transverse momenta:

\begin{equation}
T^{\mathrm{T}} = \mathrm{max}_{\hat{n}}\frac{\sum_{\mathrm i} 
\left|\vec{p}_{\mathrm T_{\mathrm i}} \cdot 
\hat{n}\right|}{\sum_{\mathrm i} \left|\vec{p}_{\mathrm T_{\mathrm i}}\right|}
\label{eq:Tht}
\end{equation}

Transverse thrust ranges from $T^{\mathrm T}$=$1$ to $T^{\mathrm T}$=$2/\pi$
($\langle|\cos\theta|\rangle$) for a back-to-back and an isotropic
distribution of particles in the transverse plane, respectively.
To minimize systematics associated with the busy environment in a 
$p\overline{p}$ collider, we use only 
the two leading jets in the event, reconstructed with a
$k_{\perp}$ $D$=$1$ algorithm, rather than using all calorimeter towers.
Other particles in the event are inferred from the angular distribution of
the two leading jets.

The measurement of the dijet transverse thrust, $T^{\mathrm T}_2$ presents
a good opportunity to test resummation models, as well as
the recently developed NLO pQCD three-jet generators~\cite{trirad,nagy}.
$T^{\mathrm T}_2$ is binned in terms of $H_{\mathrm{T3}}$,
defined as the scalar sum of the transverse momenta of the three leading 
jets of the event $p_{\mathrm{T1}}+p_{\mathrm{T2}}+p_{\mathrm{T3}}$.
$H_{\mathrm{T3}}$ is an estimator of the energy scale
of the event. 
The lowest order at which pQCD does not give the trivial result 
$T_2^{\mathrm T}$=$1$
is $O(\alpha_{\mathrm s}^{3})$, corresponding to up to three parton jets in the
final state. A
$O(\alpha_{\mathrm s}^{3})$ calculation, like {\sc jetrad}, 
does not cover the whole physical range of $T_2^{\mathrm T}$. 
A $O(\alpha_{\mathrm s}^{3})$ 
prediction will also fail at $T_2^{\mathrm T}\rightarrow 1$, where soft 
radiation in
dijet events contribute large logarithms which need to be resummed.

\begin{figure}[htbp]
\vskip-5mm
\includegraphics[width=8cm]{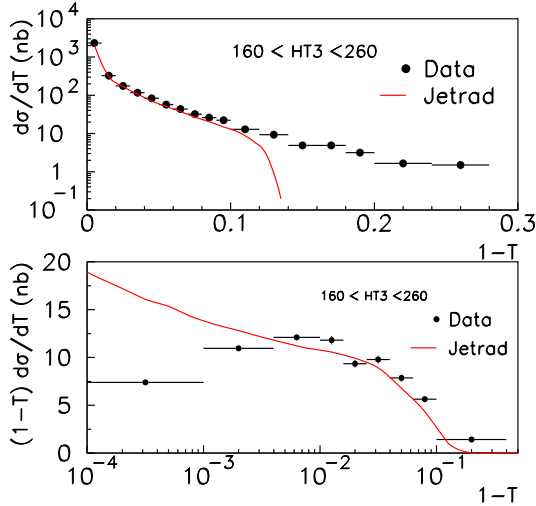}
\vskip-1cm
\caption{Thrust cross section
as a function of $1-T_2^{\mathrm T}$ for the lowest
$H_{\mathrm{T3}}$ bin: 160-260~GeV.}
\label{fig:thrust1}
\vskip-1cm
\end{figure}

\begin{figure}[htbp]
\vskip-5mm
\includegraphics[width=8cm]{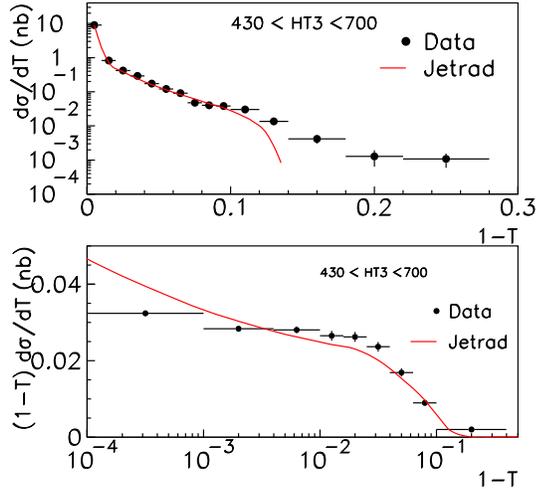}
\vskip-1cm
\caption{Thrust cross section 
as a function of $1-T_2^{\mathrm T}$ for the highest 
$H_{\mathrm{T3}}$ bin: 430-700~GeV.}
\label{fig:thrust2}
\end{figure}

Figures~\ref{fig:thrust1} and~\ref{fig:thrust2} show thrust cross sections 
as a function of $1-T_2^{\mathrm T}$ for the lowest and highest 
$H_{\mathrm{T3}}$ bins: 160-260~GeV and 
430-700~GeV. 

The error bars are statistical, and the systematic 
uncertainties are in the 15-30$\%$ range. 
The higher the value of $1-T_2^{\mathrm T}$, the more dominant is the 
contribution of high order terms in $\alpha_{\mathrm s}$. This 
explains the disagreement between the 
data and the $\alpha_{\mathrm s}^{3}$ prediction {\sc jetrad} in the
high $1-T_2^{\mathrm T}$ range. 
For example, for $T_2^{\mathrm T}$ between $\sqrt{2}/2$ and  
$\sqrt{3}/2$, the $\alpha_{\mathrm s}^{4}$ terms contribute to LO. 
When we expand the range of thrust around unity using a logarithmic scale, 
we verify that {\sc jetrad} also fails, suggesting the need to
resum higher order terms to improve agreement.
Table~\ref{tab:thrustchi} displays the agreement probability between 
the D\O\ data and {\sc jetrad}, which decreases sharply as we incorporate
data points in the high and low $1-T_2^{\mathrm T}$ range. The probabilities 
are calculated using the full covariance error matrix in a $\chi^2$ test.

\begin{table}[htb]
\caption{Agreement probability between 
the D\O\ data and {\sc jetrad},
using the full covariance error matrix in a $\chi^2$ test.}
\label{tab:thrustchi}
\newcommand{\m}{\hphantom{$-$}}
\newcommand{\cc}[1]{\multicolumn{1}{c}{#1}}
\renewcommand{\arraystretch}{1.2} 
\begin{tabular}{@{}llll}
\hline
$1-T_2^{\mathrm T}$ Range & \cc{$\chi^2$} & $\#$ d.o.f. & Prob.($\%$) \\
\hline
$0-0.1$    & \m10 & \m10 & \m42 \\
$0-0.12$   & \m13 & \m11 & \m30 \\
$0-0.14$   & \m42 & \m12 & \m0.004 \\
\hline
$10^{-2.4}-0.063$  & \m2.7  & \m5 & \m75 \\
$10^{-3}-0.063 $   & \m3.8  & \m6 & \m71 \\
$10^{-4}-0.063$    & \m95   & \m7 & \m0 \\
\hline
\end{tabular}\\[2pt]
\end{table}

\section{Conclusions}

D\O\ has successfully implemented and calibrated a $k_{\perp}$
jet algorithm in a $p\overline{p}$ collider. Quark and gluon
jets have a different structure consistent with the {\sc herwig} prediction and
previous experimental results from $e^{+}-e^{-}$ colliders.
The thrust cross section measurements indicate the need to include 
higher than $\alpha_{\mathrm s}^{3}$ terms and resumation in the theoretical
predictions. They also offer an excellent opportunity to test the 
recently developed NLO three jet generators~\cite{trirad,nagy}.  
The marginal agreement between the measurement of the 
particle level inclusive $k_{\perp}$ jet cross section with the 
$\alpha_{\mathrm s}^{3}$ theory is opening a debate on matters 
such us hadronization, underlying event, and algorithm definition.

%
%

%
%



\end{document}